\documentclass[aps,prl,showpacs,twocolumn]{revtex4}

\usepackage{graphicx}
\usepackage[ansinew]{inputenc}
\usepackage{color}
\usepackage{float}
\definecolor{Blue}{rgb}{0.3,0.3,0.9}
\begin{document}

\title{Plasmaron excitations in p(2$\times$2)-K/Graphite}
\author{V. Chis$^{1}$}
\email{vasse.chis@yahoo.com}
\author{V.M. Silkin$^{1,2,3}$}
\author{B. Hellsing$^{4,1}$}

\affiliation{$^1$Donostia International Physics Center (DIPC), 20018 San Sebasti\'{a}n, Spain \\
$^2$Depto. de F\'{\i}sica de Materiales and Centro
Mixto CSIC--UPV/EHU, Facultad de Ciencias Qu\'{\i}micas, Universidad
del Pa\'{\i}s Vasco, Apdo. 1072, 20080 San Sebasti\'an,  Spain\\
$^3$IKERBASQUE, Basque Foundation for Science, 48011, Bilbao, Spain \\
$^4$Department of Physics, Gothenburg University, S-41296 Gothenburg, Sweden}

\begin{abstract}
A new type of plasmarons formed by the compound of photoelectrons and acoustic surface plasmon excitations is investigated in the system p(2$\times$ 2)-K/Graphite. The physics behind these types of plasmarons, $e$-plasmarons, is different from the ones recently found in graphene where the loss feature is argued to result from the photohole-plasmon interaction, $h$-plasmarons. Based on $first$ $principles$ methods we calculate the dispersion of $e$-plasmaron excitation rate which yields a broad feature below the parabolic quantum-well band with a peak about 0.4 eV below the quantum well band in the $\bar{\Gamma}$-point.
\end{abstract}


\maketitle

\section{Introduction}

In photonics light is used for information processing. The dimension of channels directing the light is limited by its wave length, which is typically several hundred nanometers. However, if the information could be transferred to plasmons, these could in turn be guided through structures measured on the nanometer scale, as the plasmon wave length is of the order nanometers. This is the basis of the plasmonic devices that merge photonics and electronics \cite{Raether_80,Raether_88,teb_98,eoz_06}. The transformation of information from a photon to a plasmon requires some intermediate electronic excitation as the photon carries a negligible momentum. These intermediate excitations could be elementary charge excitations, electrons or holes.

Angular Resolved Photoemission Spectroscopy (ARPES) is a suitable tool to obtain information about the efficiency of the link between the incoming photon and the excitation of plasmons. However, the outgoing photoelectron carries information, not only about the $\it{photohole-plasmon}$ interaction described by the spectral function, but also about the direct $\it{photoelectron-plasmon}$ interaction. We report on the importance of the photoelectron-acoustic surface plasmon interaction yielding what we denote as $e$-plasmarons, not previously considered.

The importance of ARPES is of tremendous importance for the understanding of properties of solids. Not only as a tool to map the single-electron band structure of compounds but also to learn about the influence of various many-body effects such as electron-electron interaction of itinerant electrons \cite{Bohm_PR53,ecbessr04,Chulkov_SS01,Borisov_PRL01}, local electron-electron Coulomb repulsion in strongly correlated materials \cite{Imada_RMP98}, the electron-phonon interaction \cite{Grimvall_81,Hellsing_JPCM02} and electron-plasmon interaction exciting so-called plasmarons \cite{Bohm_PR53,Lundqvist_PKM67}. All these many-body effects renormalize the band structure observed in ARPES.

The photoemission process can by divided in three steps, the optical excitation, propagation of the photoelectron to the surface and finally its escape into the vacuum. In the first step the spectral function A(${\bf k},\omega$), describing the photo-hole relaxation, is in general considered to produce the bulk part of the photoemission spectrum. This {\it sudden approximation} is not appropriate in many cases. The second step is known to yield surface effects leading to misinterpretations of bulk properties \cite{Maiti_EL_2001,liprl03,sefuprl04}.

The features appearing in the photoemission spectra at binding energies higher than the single-particle bands could reflect {\it extrinsic} loss processes taking place in steps two and three.  However, this part of the spectrum could also reflect {\it intrinsic} processes, which should be taken care of by the spectral function, giving rise to, e.g. lower-Hubbard bands for strongly correlated materials. As pointed out by Guzzo {\it et al.} \cite{Guzzo_EPJ_2012} the analysis and understanding of this part of the spectrum is poor in valence-band spectroscopy.

A recent study of graphene, comparing ARPES data and calculations of the spectral function focusing on the dynamics of the electron-electron interaction, suggests presence of plasmarons \cite{Bostwick_SCIENCE_2010}. The plasmaron, a composite excitation due to the long range interaction between elementary charges (electrons or holes) and plasmons was suggested almost half a century ago \cite{Hedin_SSC_1967,Lundqvist_PKM67}. The interpretation of the ARPES data \cite{Bostwick_SCIENCE_2010} is that the photoelectrons reaching the detector carries information of the interaction between the hole left behind and plasmons. We denote this excitations $h$-plasmarons, where $h$ refers to the hole-plasmon interaction. In a recent tunneling spectroscopy study plasmaron excitations of the hole-type is claimed to be observed also for two-dimensional (2D) quantum well systems by Dial {\it et al.} \cite{Dial_PRB_2012}. A recent theoretical study of bilayer graphene, calculating the spectral function taking into account the electron-electron interaction within the Random-phase approximation, reveals plasmaron excitations \cite{Sensarma_PRB_2011}. The excitations are in this case due to the photohole-plasmon interaction and thus of the type $h$-plasmarons. These observations together indicate that plasmaron are a general feature of quasi 2D electron systems.

There is a possibility that in step three of the photoemission process, the unscreened photoelectron, when detached from the electron gas of the compound on its way to the spectrometer, but still close enough to induce excitations in the material, will create collective excitations and thus loose some of its energy and momentum. This phenomena should be in particular important in cases when low energy surface localized plasmons could be excited. Thus it is in principle possible that the additional lower energy feature in the ARPES spectrum is due to plasmon excitations in the solid due to the electromagnetic field created by the decoupled outgoing photoelectron. In this case the elementary particle is an electron and not a hole and we denote the composite "particle" an $e$-plasmaron. The picture is obviously different from the interpretation that the feature reflects $h$-plasmaron excitations in the solid.

In the present study we focus on the third step of the photoemission process for the system p(2$\times$2)-K/Graphite and calculate the dispersion of the $e$-plasmaron excitation rate. The system is ideal for the study of the third step due to the quasi-2D character of the electron structure which rule out the importance of the second step. The results indicate that losses due to the interaction of the escaping photoelectron and surface localized plasmons could give rise to satellites in ARPES spectrum. The ARPES data by Algdal {\it et al.} \cite{Algdal_PRB_2006} for the system p(2$\times$2)-K/Graphite show an asymmetry of the  quantum well peak near normal emission which is consistent with additional losses below the peak.

For systems with an electron surface state band crossing the Fermi level, a plasmon localized at the surface and characterized by a sound-like dispersion, so-called Acoustic Surface Plasmon (ASP), has been predicted to exist \cite{vsi_04,vsi_05,Pitarke_RPP_2007,vsi_10}. Later on Electron Energy Loss Spectroscopy experiments have confirmed the presence of the ASP mode at the Be(0001) surface in good agreement with {\it first principles} calculations \cite{bdi_07} and noble metal surfaces \cite{spa_10,kpo_10,jamuprb12,vasmprl13,piwejpcc13} and graphene adsorbed on metal substrate \cite{lafonjp11,pomaprb11,pomaprb12}. In cases when surface localized quantum well states are formed, e.g. when atomic layers of alkali metals are adsorbed on a metal surface the possibility opens up to design ASP by varying the depth (type of alkali atoms) and the width (number of layers) of the quantum well.

\section{Theory}

The energy loss induced by the escaping photoelectron in an ARPES experiment is given by the
rate of electronic surface excitations. In first order time dependent perturbation theory, "golden-rule", we have that the rate of $e$-plasmaron excitation is given by \cite{Liebsch_97}.

\begin{eqnarray}\
W(\omega) &=&  2 \pi \sum_{{\bf k}_{\|}, {\bf k}_{\|}',n,n'}
(f_{n{\bf k}_{\|}}-f_{n'{\bf k}_{\|}'})|\langle\psi_{n'{\bf k}'_{\|}}^{*}|\phi_{scf}|\psi_{n{\bf k}_{\|}}\rangle|^{2}  \nonumber \\
&\times& \delta(\epsilon_{n'{\bf k}'_{\|}}-\epsilon_{n{\bf k}_{\|}}-\hbar \omega)  ,
\label{eq:GR}
\end{eqnarray}
where $f$ is the Femi-Dirac factors, $\psi_{n{\bf k}_{\|}}$ the one-electron wave functions with the corresponding energies $\epsilon_{n{\bf k}_{\|}}$ and $\phi_{scf}$ is the self-consistent potential caused by the photo electron which drives the single electron excitations from state $n{\bf k}_{\|}$ to state $n{\bf k}_{\|}'$.
Equation (\ref{eq:GR}) can be expressed in terms of the bare density-density response function $\chi_{0}({\bf r},{\bf r}',\omega)$ evaluated in a slab geometry \cite{sichprl04}
\begin{eqnarray}\
\label{eq:xhi}
\chi_{0}({\bf r},{\bf r'},\omega) &=& \sum_{{\bf k},{\bf k}',n,n'} (f_{n{\bf k}_{\|}}-f_{n'{\bf k}_{\|}'}) \\
&\times&
\frac{ \psi_{{\bf k}n}^*({\bf r}) \psi_{{\bf k}'n'}({\bf r})
\psi_{{\bf k}'n'}^*({\bf r}') \psi_{{\bf k}n}({\bf r}')}
{\epsilon_{{\bf k}'n'}-\epsilon_{{\bf k}n}-\hbar \omega + i\delta} \ \nonumber.
\end{eqnarray}
Then we have
\begin{eqnarray}
\label{eq:GR1}
W(\omega)= -2 \ \ \ \ \ \ \ \ \ \ \ \ \ \ \ \ \ \ \ \ \ \ \ \ \ \ \ \ \ \ \ \ \ \ \ \ \ \ \ \ \ \ \ \ \ \ \ \ \ \ \  \\
 \times {\rm Im} \left[\int {\rm d}{\bf r} \int {\rm d}{\bf r}_{1} \chi_{0}({\bf r},{\bf r}_{1},\omega)\phi_{scf}({\bf r},\omega)\phi_{scf}^{*}({\bf r}_{1},\omega) \right] \ \nonumber ,
\end{eqnarray}
where ${\rm Im}$ represents the imaginary part. Note that we use the notation ${\bf r} \equiv ({\bf r}_{\|},z)$ and ${\bf k} \equiv ({\bf k}_{\|},k_{z})$ for vectors in real and reciprocal spaces, respectively. Within a linear response theory we have in momentum and frequency space
\begin{eqnarray}\
\label{eq:rel}
\chi_{0} \phi_{scf} &=& \chi_{0} \phi_{ext} + \chi_{0} \delta \phi = \ \nonumber \\
 & & \chi_{0} \phi_{ext} + \chi_{0} K \delta n = \nonumber \\
 & & (\chi_{0} + \chi_{0} K \chi)\phi_{ext} = \nonumber \\
 & & \chi \phi_{ext} = \delta n \ ,
\end{eqnarray}
where $\phi_{ext}$ is the external potential due to the photo electron, $\delta \phi$ the induced potential caused by the induced electron density $\delta n$. The kernel $K$ in real space is explicitly given by
\begin{eqnarray}
K({\bf r},{\bf r}') =
\frac{1}{|{\bf r}-{\bf r}'|} + V'_{xc}[n_{0}({\bf r})]\delta({\bf r}-{\bf r}') \ ,
\end{eqnarray}
where $V_{xc}$ is the exchange-correlation potential. According to Eq. (\ref{eq:rel}) the full response function is determined self-consistently by the Dyson equation
\begin{eqnarray}
\chi = \chi_{0} + \chi_{0} K \chi \ .
\end{eqnarray}
The induced density $\delta n$ is given by
\begin{eqnarray}
\delta n({\bf r},\omega) &=& \int {\rm d}{\bf r}_1 \chi({\bf r},{\bf r}_1,\omega)\phi_{ext}({\bf r}_1,\omega) \ \\ \nonumber
 &=&\int {\rm d}{\bf r}_1 \chi_{0}({\bf r},{\bf r}_1,\omega)\phi_{scf}({\bf r}_1,\omega).
\end{eqnarray}

We can thus write
\begin{eqnarray}\
\label{eq:GR3}
W(\omega) &=& -2 \,{\rm Im}  \left[\int {\rm d}{\bf r} \
\phi_{scf}^{*}({\bf r},\omega) \delta n({\bf r},\omega)\right] \nonumber \\
&=&-2 \,{\rm Im}  \left[\int {\rm d}{\bf r} \phi_{ext}^{*}({\bf r},\omega) \delta n({\bf r},\omega)\right] \ .
\end{eqnarray}
We consider an ARPES experiment and the possibility that an ejected photoelectron will lose part of its energy somewhere in between the surface and the detector at an average distance $z_{0}$ from the surface. We keep $z_{0}$ as a parameter to vary later on. We assume that the distance from the surface is far enough that the "spill-out" electron density from the surface is neglectible at $z=z_{0}$ . The external potential must then fulfill Laplace equation \cite{Liebsch_97}
\begin{eqnarray}\
\label{eq:ext}
\phi_{ext}({\bf r}_{\|},z,\omega) =
-\frac{1}{A}\sum_{{\bf q}_{\|}}\frac{2\pi}{q_{\|}}\,e^{q_{\|}(z-z_{0})}e^{{\rm i}({\bf q}_{\|}\cdot{\bf r}_{\|}-\omega t)} ,
\end{eqnarray}
where $A$ is the surface area and $q_{\|}=|{\bf q}_{\|}|$. The rate of energy loss can then be expressed in terms of the surface response function \( g({\bf q}_{\|},\omega) \) \cite{Persson_PRB_1984}
\begin{eqnarray}\
\label{eq:g} g({\bf q}_{\|},\omega) =
\int {\rm d}z \ e^{q_{\|}z} \delta n(z,{\bf q}_{\|},\omega)
\label{eq_g}
\end{eqnarray}
and accordingly
\begin{eqnarray}\
\label{eq:wfin}
W(\omega) =
 \frac{4\pi}{A} \sum_{{\bf q}_{\|}}\frac{e^{-q_{\|}z_{0}}}{q_{\|}} \,{\rm Im}[g({\bf q_{\|}},\omega)].
\end{eqnarray}
%
\section{Calculations}

\begin{figure}[H]
\includegraphics[width=85 mm]{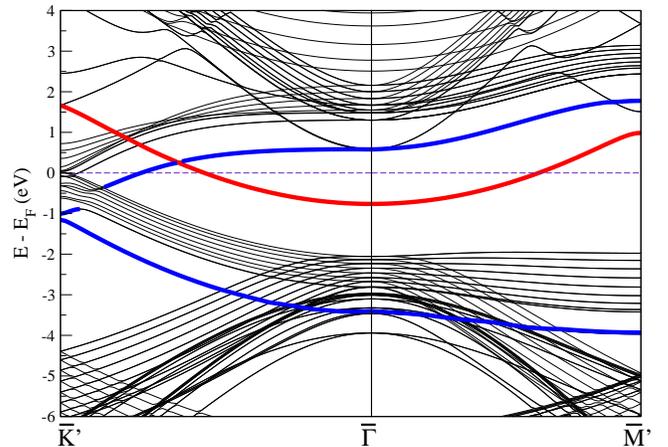}
\caption{(Color online) Calculated p(2$\times$2)-K/Graphite band structure \cite{Chis_PRB_2011}. Red and blue color lines indicate the quantum well band and the lowest and the highest branches of the folded $\pi^{*}$ and $\pi$ bands, respectively. $\bar{\text{K}'}$ and $\bar{\text{M}'}$, represent the $\bar{\text{K}}$ and $\bar{\text{M}}$ points of the folded band structure due to the (2$\times$2) overlayer of potassium.}
\label{fig_bands}
\end{figure}

With this theoretical background we proceed to the specific system, a monolayer of potassium
on graphite, p(2$\times$ 2)-K/Graphite. According to the $first$ $principles$ calculations by
Chis {\it et al.} \cite{Chis_PRB_2011} $two$ quasi-2D electron systems on top of each other is formed.
At the top, a quantum well (QW) 2D system is formed with an energy band centered at the $\bar{\Gamma}$-point of the Brillouin zone (BZ) (see red colored line in Fig. \ref{fig_bands}). Another quasi-2D system is formed below the QW system and essentially located to the uppermost single graphite layer. The band structure localized to this doped graphene-type of layer has an anti-bonding and a bonding character. The lower branch of the down-shifted anti-bonding $\pi^{*}$ band  (folded due to the (2$\times$2) potassium overlayer) and the upper branch of the folded bonding $\pi$ band are marked by blue color in Fig. \ref{fig_bands}. The bands meet at the $\bar{\text{K}'}$-point of the BZ, about 1 eV below the Fermi level.
\subsection{Acoustic surface plasmons}
\begin{figure}
\includegraphics[width=90 mm]{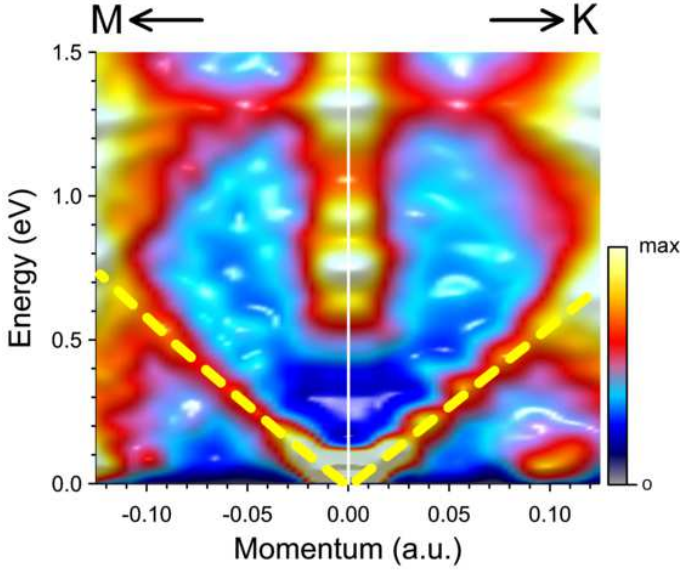}
\caption{(Color online) The calculated normalized surface loss function, Im$[g({\bf q}_{\|},\omega)/(q_{\|}\omega)]$, along the  $\bar{\Gamma}$-$\bar{\text{M}}$ and $\bar{\Gamma}$-$\bar{\text{K}}$ directions. Dashed lines highlight the sound-like dispersion of the ASP. The features at higher energies correspond to interband transition not relevant for this work.}
\label{fig_ASP_dispersion}
\end{figure}

We proceed and calculate the linear response spectrum caused by the escaping photoelectron, still close to the surface but detached from the surface electron density. In order to find out about the existence of ASPs we calculated the surface loss function, Im$[g]$ within the Time Dependent Density Functional Theory scheme \cite{Petersilka_PRL96} with one-electron wave functions and corresponding eigen energies from the band structure calculation discussed above. The result of the calculation is shown in Fig. \ref{fig_ASP_dispersion}

Indeed we find the ASP in the energy range 0-0.7 eV with a momentum transfer span up to about 0.12 a.u. The interpretation is that for $q_{\|}$ $>$ 0.12 a.u. the coherence of single electron excitations forming the collective plasmon excitation is lost due to incoherent electron-hole pair excitations. In Fig. \ref{fig_ASP_dispersion} we notice that the ASP dispersion along the $\bar{\Gamma}$-$\bar{\text{M}}$ and $\bar{\Gamma}$-$\bar{\text{K}}$ directions are very similar. Based on this observation we will further on assume that the ASP dispersion is isotropic in the (x,y)-plane.

Having $two$ 2D systems on top of each other could in principle lead to two branches in the ASP dispersion. However, there is clearly only one ASP branch with a group velocity of $c$ = 0.24 a.u. at small $q_{\|}$ and reducing down to 0.22 a.u. at higher $q_{\|}$. The initial velocity $c$ should according to Pitarke {\it et al.} \cite{pinaprb04}, be set by the Fermi velocity of the 2D carriers. This is consistent with the band structure in Fig. \ref{fig_bands}. The slope of the bands when crossing the Fermi level, moving away from the $\bar{\Gamma}$-point for the QW system and away from the $\bar{K}$-point for the "graphene" system happens to be very similar, $v_{F}$ $\approx$ 0.23 a.u.

\subsection{$e$-plasmaron dispersion}
A photo-excited electron with initial parallel wave vector ${\bf k}'_{\|}$ will with some probability be inelastically scattered to ${\bf k}_{\|}$ while exciting an ASP with momentum ${\bf q}_{\|}={\bf k}_{\|}-{\bf k}'_{\|}$. Thus the yield of photoelectrons with momentum ${\bf k}_{\|}$ will have a main peak, corresponding to the electrons with initial momentum ${\bf k}_{\|}$ having absorbed fully the photon energy and with a broadening due to the finite lifetime of the photo-hole left behind. In addition a satellite structure will appear at higher binding energies due to scattering from all ${\bf k}'_{\|}$ satisfying  ${\bf k}'_{\|}={\bf k}_{\|}-{\bf q}_{\|}$, having excited an ASP with momentum ${\bf q}_{\|}$.

In the energy range of the ASP, 0-0.5 eV, the electron group velocity scanned over the entire BZ indicates that most contribution to the imaginary part of dielectric function comes from the QW band, which reflects the fact that the density of states of the quasi-2D electron gas of the QW is fairly independent of energy, while for the "graphene" layer it decreases approximately linear when approaching the $\bar{\text{K}}$-point. This means that near the Fermi level the  electron density of states will be comparatively low in the "graphene" layer.

Based on this we will in this study only consider the $e$-plasmaron excitations due to the coupling between the photoelectron, originating from the QW band with the ASP. We calculate the ${\bf k}_{\|}$-resolved photoelectron energy loss due to ASP excitations, which is equivalent to the dispersion of the $e$-plasmaron excitations. This can be carried out from the expression given in Eq. (\ref{eq:wfin}). The QW band structure shown in Fig. \ref{fig_bands} reveals that the dispersion along the $\bar{\Gamma}$-$\bar{\text{M}'}$ and $\bar{\Gamma}$-$\bar{\text{K}'}$ directions are similar. Thus we will assume that the QW band is isotropic in the surface plane, just as we previously concluded considering the ASP dispersion.

\begin{figure}
\includegraphics[width=65 mm,clip=true]{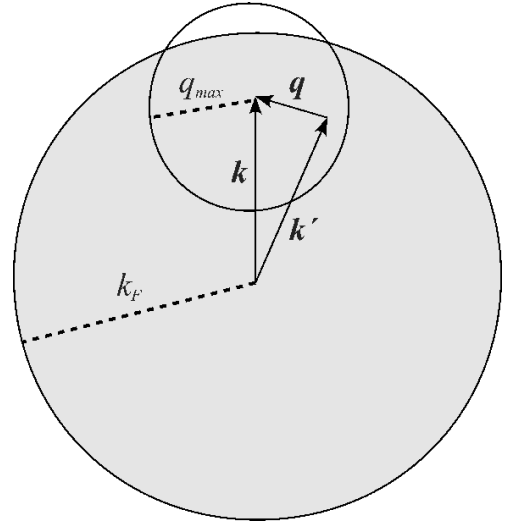}
\caption{Illustration of the integration performed in Eq. (\ref{eq_Wkomega}). The grey area represents the occupied part of the 2D electron states with the Fermi momentum radius $k_{F}$. For any fixed vector $\bf{k}$ $\equiv$ ${\bf k}_{\|}$ we integrate over all occupied states $\it{\bf{k}'}$ $\equiv$ ${\bf k}'_{\|}$ enclosed in the circular disk with radius given by the maximum absolute magnitude of the ASP wave vector $q_{max}$.}
\label{fig_integration}
\end{figure}

For a fixed vector ${\bf k}_{\|}$ we integrate over all occupied states ${\bf k}'_{\|}$ (grey area in Fig. \ref{fig_integration}) enclosed in the circular disk with a radius given by $q_{max}$ (illustrated in Fig. \ref{fig_integration}).
\begin{eqnarray}
W(k_{\|},\epsilon) &=& \frac{2}{\pi}\int_{0}^{k_{F}}{\rm d}k'_{\|} \int_{0}^{2\pi}{\rm d}\alpha \  \frac{k'_{\|}}{q_{\|}} \ e^{-q_{\|}z_0} \Theta(q_{\rm max}-q_{\|}) \   \nonumber \\
&\times& {\rm Im}[g(q_{\|},\epsilon-\epsilon_{b}+\hbar\omega(k'_{\|}))]   \ ,
\label{eq_Wkomega}
\end{eqnarray}
where  $k_{F}$ is the band Fermi wave vector, $k_{\|}=|{\bf k}_{\|}|$, $k'_{\|}=|{\bf k}'_{\|}|$, $\alpha$ the angle between the vectors ${\bf k}_{\|}$ and ${\bf k}'_{\|}$. $\Theta(x)$ is the Heaviside step function, $\epsilon_{b}$ the binding energy in the $\bar{\Gamma}$-point, $q_{\|}  = (k_{\|}^{2} + k_{\|}^{'2} - 2k_{\|} k'_{\|} cos \alpha)^{1/2}$, $q_{max}$ the maximum wave vector up to which the ASP dispersion is well defined and $\hbar\omega(k_{\|})$~-~$\epsilon_{b}$ is the band dispersion relative the Fermi energy.

\begin{figure}
\includegraphics[width=85 mm]{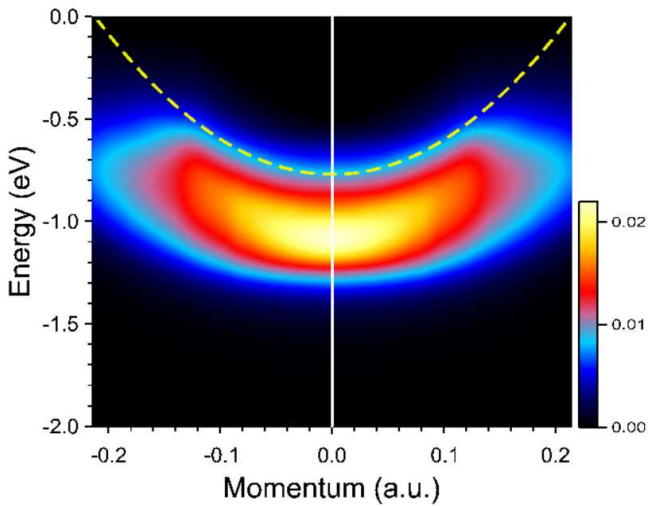}
\caption{(Color online) Plasmaron dispersion. The dashed line show the calculated dispersion of the quantum well band. The energy scale is relative to the Fermi level. The colored scale gives the rate of $e$-plasmaron excitation in atomic units (a.u.) for the case $z_{0}$=5 a.u.}
\label{fig_plasmaron}
\end{figure}

For the system we consider  $k_{F}$ = 0.23 a.u., $q_{max}$=0.1 a.u., $e_{b}$ = 0.76 eV and the QW band dispersion $\hbar\omega(k_{\|})$ = $\epsilon_{b}(k_{\|}/k_{F})^{2}$. The calculated dispersion of the $e$-plasmaron excitations is shown in Fig. \ref{fig_plasmaron}.

The loss of energy for an electron with wave vector ${\bf k}'_{\|}$, inelastically scattered due to excitation of an ASP to a state ${\bf k}_{\|}={\bf k}'_{\|}+{\bf q}_{\|}$, is determined by a straight line with the slope $c$ (ASP velocity) drawn from the QW band at ${\bf k}'_{\|}$ to ${\bf k}_{\|}$. Due to the fact that the maximum plasmon wave vector $q_{max}$ is about half the value of the $k_{F}$ the $e$-plasmaron dispersion will essentially follow the shape of the QW band. In the $\bar{\Gamma}$-point the scattering from the full circle with radius $q_{max}$ will determine the location of the maximum intensity on the energy axis. For this system with $c$ = 0.23 a.u. the maximum is at about 0.4 eV below the bottom of the QW band.

\begin{figure}
\includegraphics[width=85 mm]{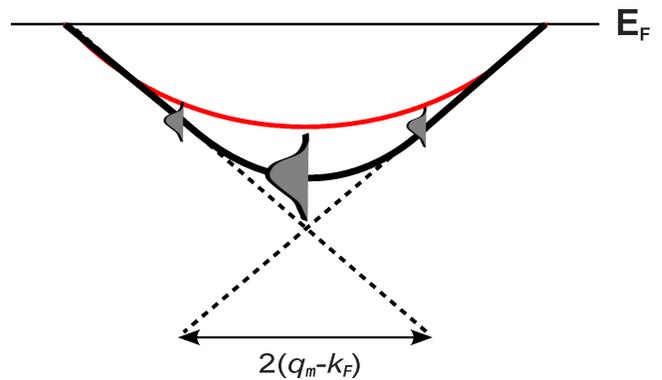}
\caption{(Color online) Schematic picture of the dispersion of the $e$-plasmaron excitation rate. The thin solid red line is the QW band, the thick black line the dispersion of the $e$-plasmaron excitation rate and the dashed black line the expected dispersion if the plasmon cut-off momentum vector \(q_{max} > k_{F}\).}
\label{fig_dispersion}
\end{figure}

It is interesting to note that the shape of the dispersion of the $e$-plasmaron excitations will depend on the maximum ASP wave vector $q_{max}$. In the case that $q_{max} > k_{F}$ the dispersion should be more or less linear. This case is illustrated by the cossing dashed lines in Fig. \ref{fig_dispersion}. Furthermore, in general, if the QW band is parabolic, the crossing point between the dashed lines should appear at an energy twice the binding energy ($\epsilon_{b}$), as the ASP velocity $c$ (slope of the dashed line) is expected to be close to $k_{F}$  \cite{vsi_04} (the slope of the QW band at the Fermi level).

\begin{figure}[h]
\includegraphics[width=85 mm,clip=true]{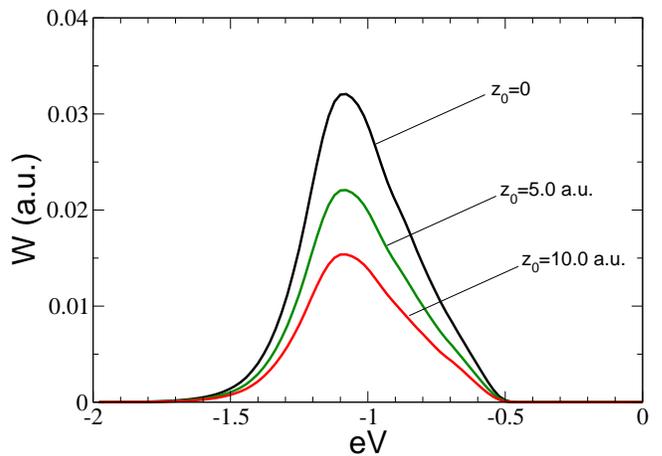}
\caption{(Color online) Plasmaron excitation rate in the $\bar{\Gamma}$-point for three different values of $z_{0}$, the distance between the surface and the position where the ASP excitation takes place.}
\label{fig_z_0}
\end{figure}

It is obvious from Eq. (\ref{eq_Wkomega}) that increasing the effective photoelectron-surface distance $z_{0}$ where the ASP excitation takes place, suppresses the higher $q_{\|}$-contributions which will reduce the excitation rate of the $e$-plasmaron. This is shown in Fig. \ref{fig_z_0} for the $\bar{\Gamma}$-point.

\section{Summary and conclusions}

Alkali monolayer adsorption on graphite forms $two$ quasi-2D systems, at the top an alkali quantum well system and below a doped graphene type of layer. Thus, we expect that for different type of alkali metals and coverages, it should be possible to design a system with two branches of Acoustic surface plasmons, each with its own velocity. In the present study of the p(2$\times$2)-K/Graphite system it turns out that only a single ASP branch exists with a velocity $c$ = 0.23 a.u. This result is consistent with the fact that the calculated band structure happens to yield almost identical Fermi velocities for the two 2D systems.

We argue that plasmarons formed due to the direct Coulomb interaction between the photoelectron and its induced electron density, creating ASP excitations should be considered as a source for loss satellite structure in ARPES for quasi-2D systems. The physics behind this type of excitation, denoted by us $e$-plasmaron, is different from that of the plasmaron recently reported for graphene \cite{Bostwick_SCIENCE_2010} and some QW systems \cite{Dial_PRB_2012}. In those cases the picture is that the photohole-plasmon interaction is responsible for the loss structure in the ARPES spectra. We denote these type of hole related plasmarons $h$-plasmarons.

In the present study of the system p(2$\times$2)-K/Graphite system, we calculate the dispersion of the $e$-plasmaron excitations and analyze in some details the character of the dispersion. Comparing with reported ARPES experiment \cite{Algdal_PRB_2006}, it might be that the plasmaron excitations are responsible for the observed high binding energy shoulder. A future challenge is to perform a comparative study of the two types of plasmarons to find out (i) which one dominates and (ii) key parameters for the $h$-plasmarons and the $e$-plasmarons, respectively.

Finally, the $e$-plasmaron excitations are important to take into account in theoretical studies of different quasi two dimensional systems since they reflect an additional channel for excitations of plasmons. This could then increase the photon-plasmon conversion yield which obviously is of interest in the field of applied plasmonics.

\section{Acknowledgements}

V. M. S.  acknowledges the partial support  from the University of the Basque Country (Grant No. IT-366-07), the
Departamento de Educaci\'on del Gobierno Vasco, the Spanish Ministerio de Ciencia e Innovaci\'on (Grant No. FIS2010-19609-C02-01)

\end{document}